\documentclass[prl,aps,twocolumn,groupedaddress,superscriptaddress,showpacs]{revtex4-1}
\usepackage{graphicx}   

\usepackage[final]{pdfpages}

\begin{document}

\title{Midair collisions enhance saltation}
\author{M. V. Carneiro}
\affiliation{Institut f\"ur Baustoffe, ETH-H\"onggerberg, Schafmattstrasse 6, 8093 Z\"urich, Switzerland}
\author {N. A. M. Ara\'ujo}
\affiliation{Institut f\"ur Baustoffe, ETH-H\"onggerberg, Schafmattstrasse 6, 8093 Z\"urich, Switzerland}
\author{T. P\"ahtz}
\affiliation{Department of Ocean Science and Engineering, Zhejiang University, 310058 Hangzhou, China}
\affiliation{State Key Laboratory of Satellite Ocean Environment Dynamics, Second Institute of Oceanography, Hangzhou, China}
\author{H. J. Herrmann}
\affiliation{Institut f\"ur Baustoffe, ETH-H\"onggerberg, Schafmattstrasse 6, 8093 Z\"urich, Switzerland}
\affiliation{Departamento de F\'isica, Universidade Federal do Cear\'a, 60451-970 Fortaleza, Cear\'a, Brazil}

\begin{abstract}
Here we address the old question in Aeolian particle transport about the role of midair collisions. 
We find that, surprisingly, these collisions do enhance the overall flux substantially. 
The effect depends strongly on restitution coefficient and wind speed. 
We can explain this observation as a consequence of a soft-bed of grains which floats above the ground and reflects the highest flying particles. 
We make the unexpected observation that the flux is maximized at an intermediate restitution coefficient of about 0.7, which is comparable to values experimentally measured for collisions between sand grains.
\end{abstract}

\maketitle

Would a sandstorm be stronger if the sand grains in air did not collide against each other? 
This question has puzzled practitioners and theoreticians alike in the past. 
Models for Aeolian sand flux~\cite{almeida,Almeida_PNAS,werner,McEwan,kok,Pahtz} become certainly much simpler if such midair collisions are neglected, but does this approximation underestimate or overestimate the value of the saturated flux? 
As opposed to experiments, the direct computer simulation of saltation, the main Aeolian transport process,  offers the possibility of switching on or off the collisions between particles or of modifying the collision parameters, such as the coefficient of restitution. 
This allows, for the first time, to precisely determine the role of midair collisions during saltation. 

\begin{figure}
\includegraphics[scale=0.16]{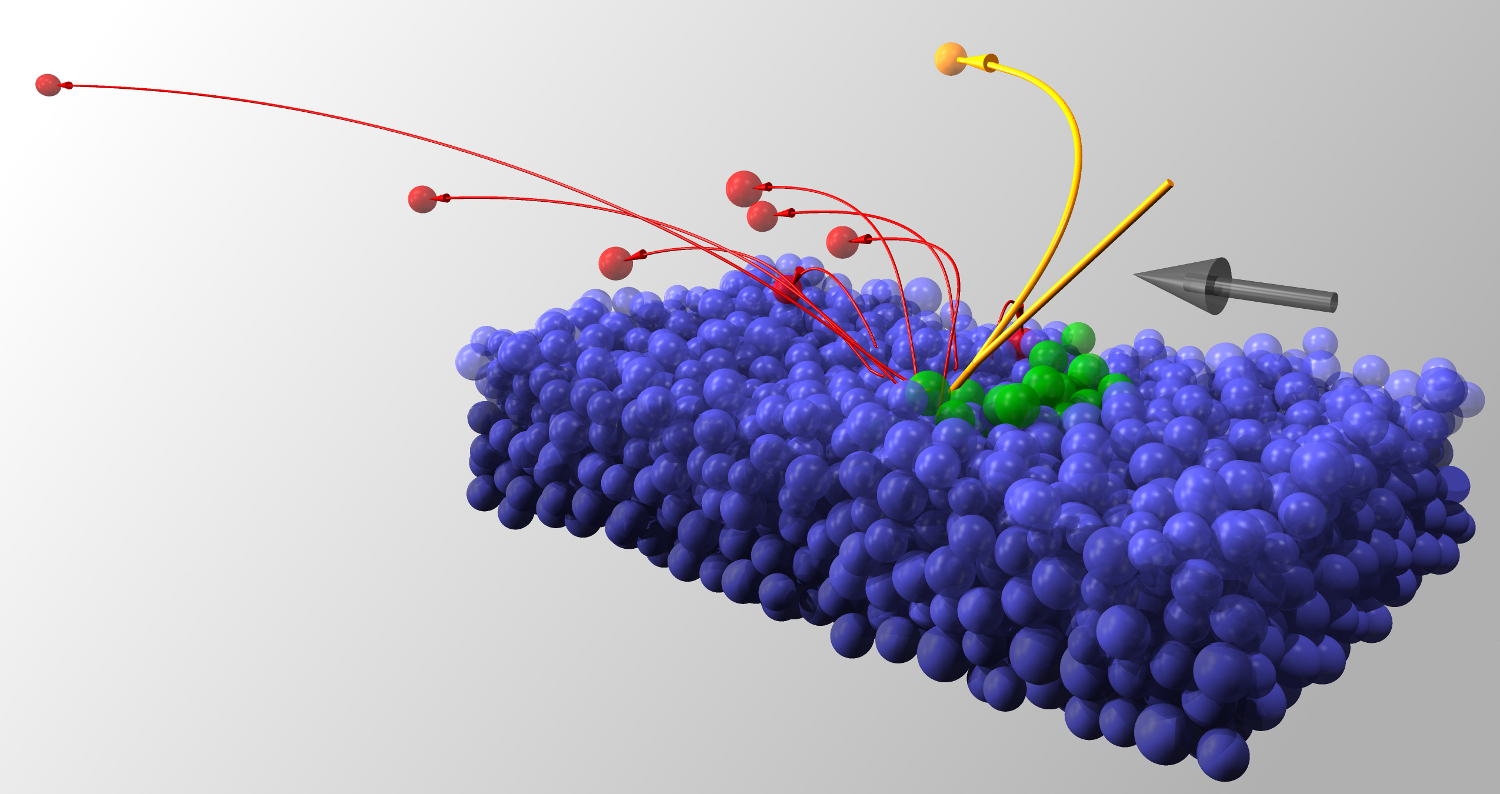}
\caption{(color online) Typical splash obtained numerically with our model in 3D. The impinging particle (yellow) bounces after ejecting other particles from the bed (red and green). While the green particles essentially only move on the ground, the red ones are lifted and dragged by the wind.}
\label{sand_bed}
\end{figure}

We discover that midair collisions are the key ingredient for understanding the relation between different concepts such as the splash~\cite{Bagnold3,kok-parteli}, the soft-bed~\cite{Williams, soresen}, and the distinction between saltons and reptons~\cite{Andreotti,Lammel}.
During saltation, particles are ejected from the granular bed in a splash, produced by the impact of fast particles, so-called saltons (yellow trajectory in Fig.~\ref{sand_bed}). 
These saltons must have the necessary kinetic energy to assure that, despite the substantial dissipation in the ground, some ejected particles can again fly sufficiently high.
After all, only high-flyers can acquire sufficient acceleration to again become saltons because the wind velocity at the ground is zero, increasing logarithmically with height. 
Our detailed study reveals the following picture: 
Due to surface irregularities, a splash produces three types of moving particles (see Fig.~\ref{sand_bed}): 
Many (green) creepers that do not leave the bed, many (red) leapers
making small jumps, remaining in regions of small wind velocities,
thereby not being able to produce a new splash, and very few saltons (yellow) which fly higher up. 
Only saltons sustain saltation. 
Both creepers and leapers (reptons) contribute considerably to the sand flux, but play a very different role in what follows. 

\begin{figure} 
\includegraphics[scale=0.17]{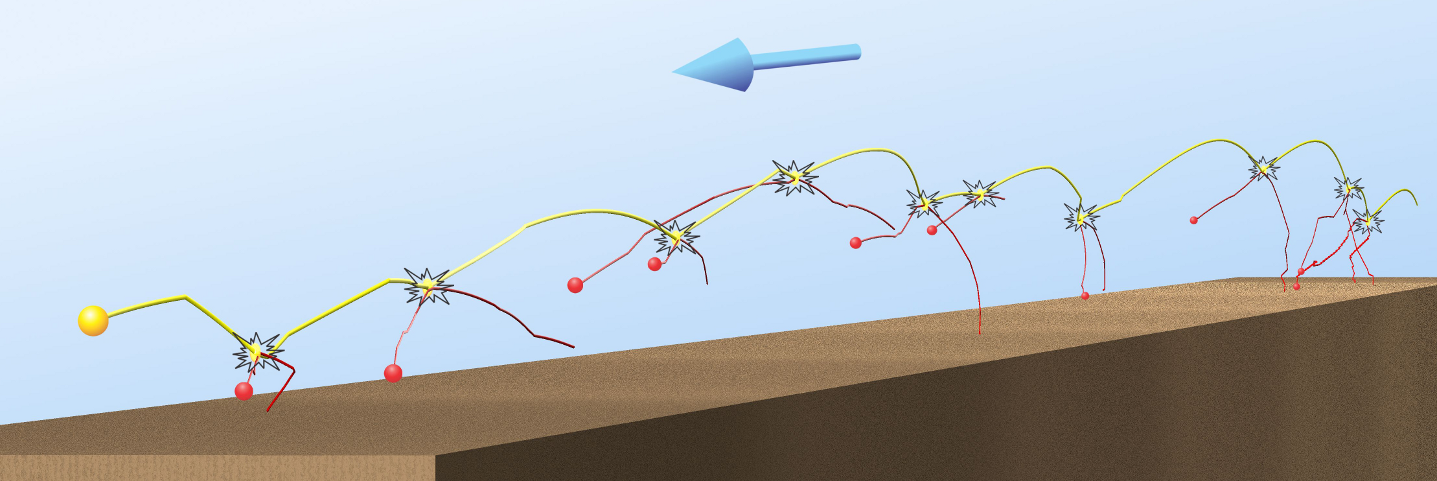}
\caption{(color online) Simulated trajectory of a salton in 3D. The (yellow, upper trajectory) salton is kept in the air by colliding against (red, lower trajectories) particles from the soft-bed. For clarity only the relevant particle trajectories are shown and the ground is schematically represented by a flat plane.}
\label{salton}
\end{figure}

Can one really make such a sharp distinction between leapers and saltons? 
After all, the collision process is random and should yield a continuous distribution of ejection velocities. 
The experimental observation of single impacts on granular packings shows a bimodal splash distribution~\cite{Mitha}, reproduced numerically by Anderson and  Haff~\cite{Anderson}, exhibiting a broad spectrum of slower particles and a small peak of faster ones. 
The analysis of these observations led Andreotti~\cite{Andreotti} to coin the terms saltons and reptons.
However, so far there is no experimental evidence that the same velocity distribution in steady state saltation is also bimodal.
What makes these two types of particles different are the midair collisions. 
Figure~\ref{salton} shows the trajectory of a typical salton simulated with the method described below in 3D:
It makes several jumps without touching the ground, rebouncing upwards each time due to a collision with a leaper. 
The probability for such a collision is reasonably high because the leapers form a rather dense layer, which is precisely the soft-bed described earlier~\cite{soresen}. 
Consequently, the salton stays longer in areas of strong wind and less time close to the ground, where the wind drag acceleration is weaker. 
This explains why the saltons acquire so much energy and can sustain the saltation process.
Summarizing, midair collisions in the soft-bed is the crucial mechanism that differentiates the saltons, and doubles the saturated flux as we will show here.

The Discrete Elements Method (DEM), which is based on particle-particle and particle-wind interactions, allows to quantitatively study the Aeolian transport~\cite{Griebel}. 
It has been used to calculate the onset of saltation and to confirm the existence of a jump in the total flux in 2D~\cite{jump}. 
Here we apply a 3D scheme to investigate the role of midair collisions.

Every sand grain is represented by a hard sphere of average diameter $D_{mean}$.
Gravity acts in vertical direction ($y$-direction) and a logarithmic wind velocity profile $u(y)$ is imposed in the horizontal direction ($x$-direction),
\begin{equation}
u(y) = \frac{u_*}{\kappa}\ln\frac{y - h_0}{y_0},
\label{wind_profile_eq}
\end{equation} 
\noindent where $y_0 = D_{mean}/30 $ is the roughness of the bed, $h_0$ the bed height, 
$\kappa = 0.4$ the von K\'arm\'an constant, and $u_*$ the wind shear velocity. The Shields number, defined as
\begin {equation}
\theta = \frac{u_*^2}{(s-1) g D_{mean}},
\end{equation}
\noindent is the pertinent dimensionless parameter that controls the wind velocity, where $s = \rho_s/\rho_w$ is the ratio between the grain and fluid density,
and $g$ the norm of the gravitational acceleration. The feedback procedure that extracts the momentum from the wind due to the acceleration of the grains is explained  in the Supplemental Material~\cite{SM}.
\begin{figure} [t]
\rotatebox{-90}{\includegraphics[scale=0.35]{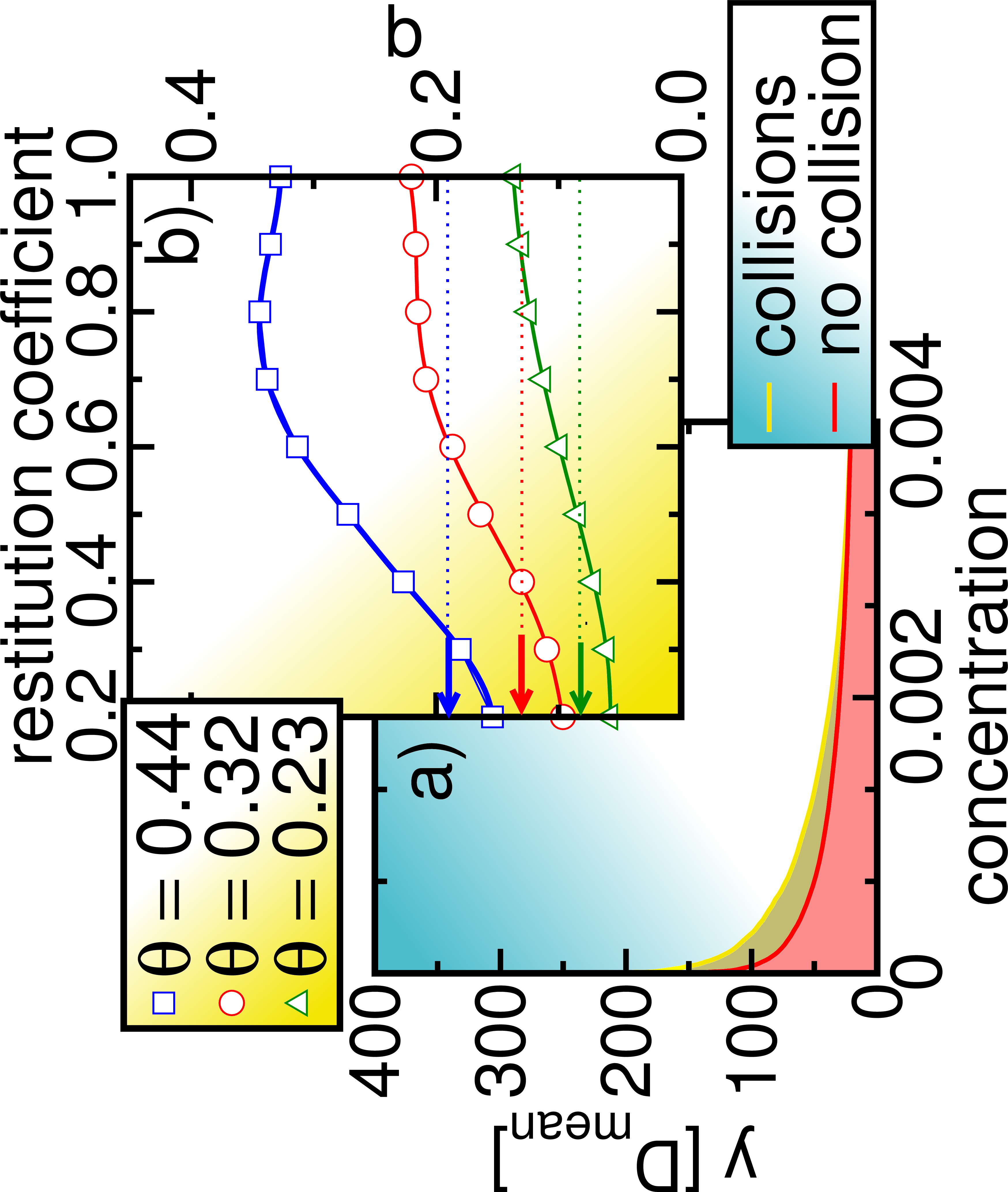}}
\caption{(color online) $(a)$ Concentration profile of particles as a function of the
height $y$ for $\theta = 0.44$ in the absence (red, lower curve) and the presence (yellow, upper curve) of midair collisions $(e = 0.7)$. $(b)$
The relation between the saturated flux and restitution coefficient,
for four different $\theta$ exhibits a peak around $e = 0.75$ for
higher shear velocities. The horizontal dashed lines show the flux without midair collisions.}
\label{concentration}
\end{figure}

We define midair collisions as those for which both particles have their center of mass above $h_0$. 
Collisions with bed-particles occur when the center of mass of at least one particle is below $h_0$. 
We study the effects of the restitution coefficient $e$ on midair collisions for a fixed restitution coefficient $e_{bed}=0.7$ for collisions with the particle bed. 
In the simulations without midair collisions, above $h_0$ the collisions are neglected, i.e., the particles are transparent to each other. 
Further information can be found in the Supplemental Material~\cite{SM}.

We consider particles of average diameter $D_{mean}~=~2~\times~10^{-4}$~m, size dispersion $\sigma_D = 0.15 D_{mean}$, and density $\rho_s=2650$ kg/m$^3$, in  a three-dimensional wind channel of dimension $(700 \times 50 \times 7.5)D^3_{mean}$ with a reflective upper boundary, placed sufficiently high to avoid any collision against it, and periodic boundaries in the other directions. 
For the fluid density, we chose $\rho_w=1.174$ kg/m$^3$.
In fact, no collision with the upper boundary has occurred in our simulations.
The lower boundary at $y = 0$, representing the deep ground,  is strongly dissipative with a fixed restitution coefficient of $e_w = 0.5$.
We consider a bed of 12 particle layers to suppress the reflection of shock waves on the lower boundary due to finite depth~\cite{rioual,rioual_thesis,rioual_2003}.
In the beginning of the simulation, a few particles are dropped at random positions to trigger saltation.

The dimensionless flux in the direction of wind is defined as
\begin{equation}
q  = \frac{1}{DA}\displaystyle\sum\limits_{i}^N m_i v_{i}^{x},
\label{dimensionless_flux}
\end{equation}
\noindent where $A=(50\times 7.5)D^2_{mean}$ is the area of the bottom of the channel, $v_{i}^{x}$ and $m_i$ are, respectively, the velocity along the  horizontal direction and the mass of the particle $i$, and $D=\rho_s\sqrt{(s-1)gD^3_{mean}}$ is a normalization constant.
The saturated flux, which is the average flux in the stationary state, does not change with the number of particles.
The granular temperature, defined as
\begin{equation}
T(y)  = \frac{1}{3N(y)}\displaystyle\sum\limits_{i}^{N(y)} m_i(\mathbf{v}_{i} - \mathbf{v}_{m}(y))^2,
\label{temperature_eq}
\end{equation}
\noindent quantifies the fluctuations around the mean velocity ${\bf v}_m(y)= 1/N(y)\sum_i^{N(y)} {\bf v}_i $. 

\begin{figure} [t]
\rotatebox{-90}{\includegraphics[scale=0.33]{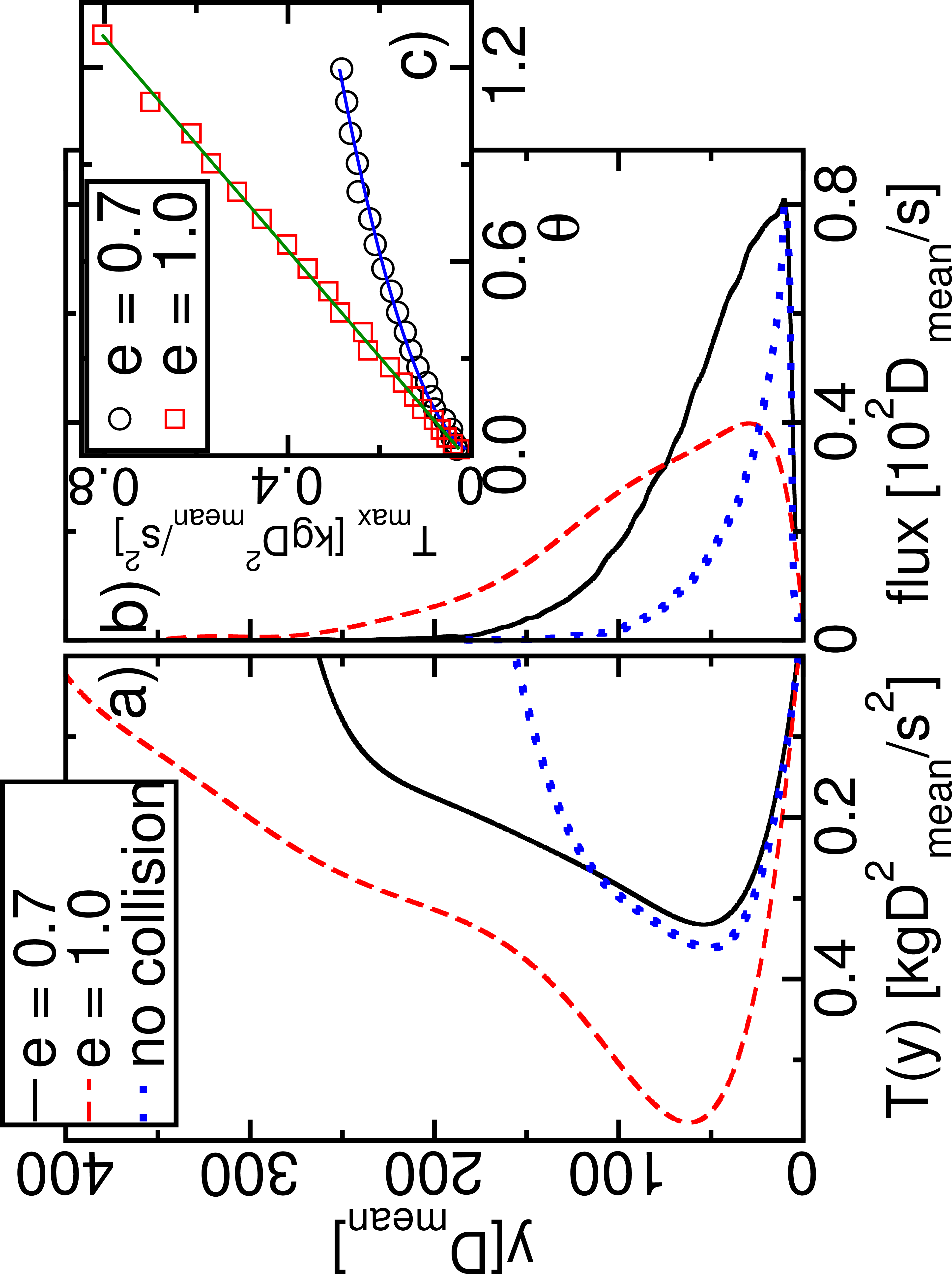}}
\caption{(color online) Temperature $(a)$ and flux $(b)$ profiles for $e=0.7$, $1.0$, and without midair collisions, for $\theta = 0.44$.  
At every height in $(a)$, the granular temperature for $e = 1.0$ is larger than $e = 0.7$. 
The flux profiles in $(b)$ confirm the higher flux for $e=0.7$.
The flux profile is defined at each height as the product of the concentration and the average velocity. 
$(c)$ Dependence of the maximum temperature on $\theta$.}
\label{temperature}
\end{figure}

The wind channel is divided along the $y$-direction in horizontal slices $(2.5 \times 50 \times 7.5)D^3_{mean}$  to calculate the profiles of particle concentration, average particle velocity, alignment, and granular temperature. 
The particle concentration is the ratio between the volume of particles and the total volume. 
The flux profile is obtained for each slice from the product of the concentration and the average particle velocity.
\begin{figure} [t]
\rotatebox{-90}{\includegraphics[scale=0.3]{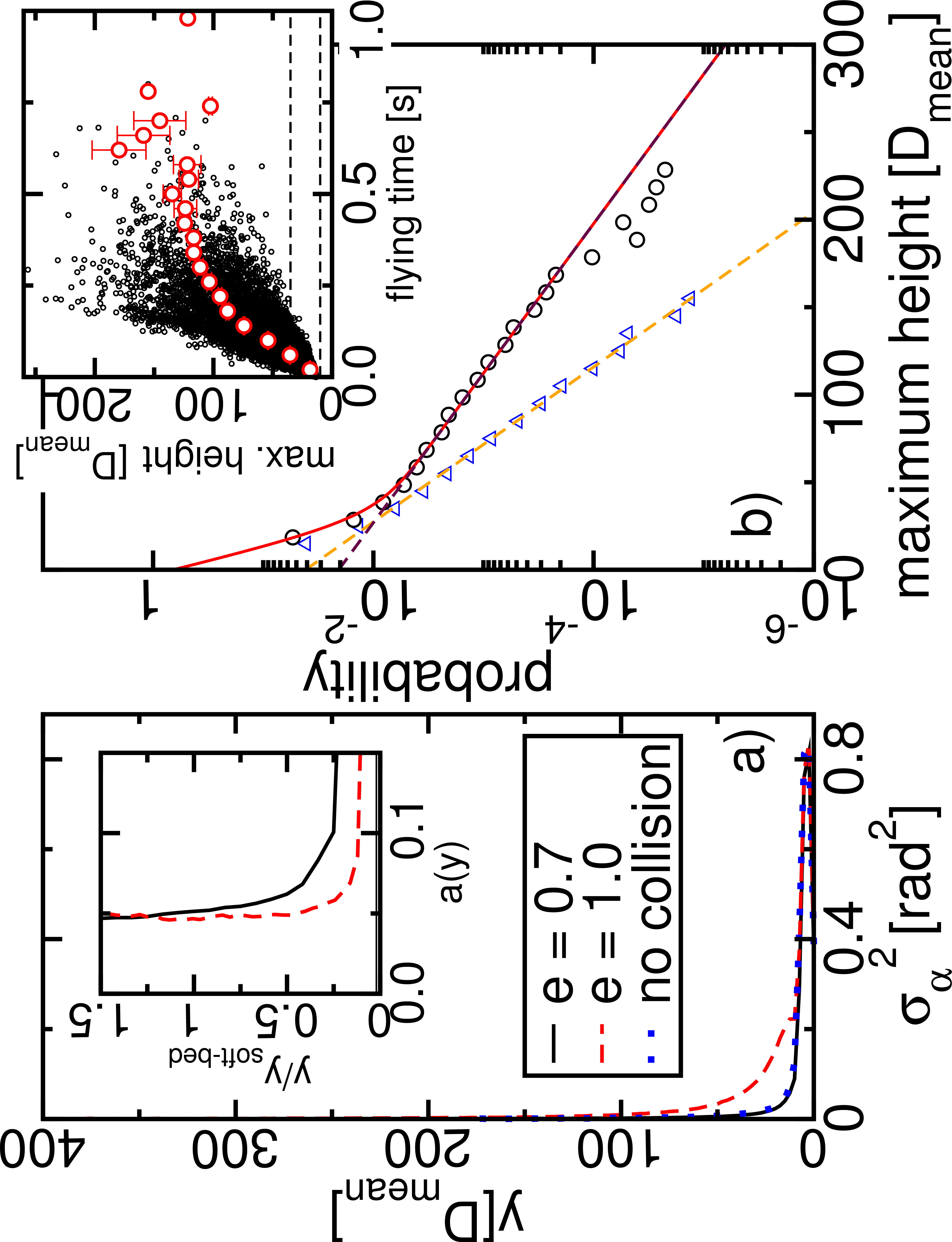}}
\caption{(color online) $(a)$ The angle variance between particle trajectories for $e= 0.7$, $e = 1.0$, and without midair collisions, for $\theta = 0.44$. 
In the inset, the profile of the inverse of the temperature anisotropy.
Also for $\theta=0.44$, $(b)$ shows the probability distribution for the maximum height, with (black circles) and without (blue triangles) midair collisions. The results in the absence of midair collisions can be fitted by the dashed (yellow) curve, given by $[0.041 exp(-0.052x)]$, and only include leapers. The case with midair collisions is fitted by the solid (red) curve which is a superposition of the contribution of leapers and the one of the saltons (dash-double-dotted (purple) curve), given by $[0.6365 exp (-0.15x) + 0.02 exp(-0.026x)]$.
In the inset, we show the relation between the flying time and the
maximum height, where the horizontal dashed line delimits the soft
bed.}
\label{angle}
\end{figure}

Figure~\ref{concentration}b shows the dependence of the saturated flux on $e$ for different $\theta$. 
For $\theta>0.44$, the saturated flux is higher for $e=0.75$ than for $e=1.0$.
The maximum at $e = 0.75$  increases substantially with $\theta$, i.e., the stronger the wind, the higher the peak.
This is also confirmed by the flux profile in Fig. 4b, where the area below the curve, which corresponds to the total flux, is larger in the presence of dissipation.
Interestingly, this optimal $e$ is comparable to the values experimentally measured for collisions between quartz grains~\cite{nasa}.
Below we will explain this maximum as the result of competition between the loss of alignment between trajectories and mounting uplift of particles with increasing $e$.

In the absence of midair collisions, all particles are reptons and follow stretched parabolic trajectories~\cite{almeida}.
Saltons emerge with midair collisions. They are located at higher positions where the wind is stronger (Fig. 2) and spend less time close to the ground, where wind drag acceleration is weaker. Thus, they are much faster and contribute much more to the overall flux.
The number of collisions and, consequently, the flying time of a salton strongly depend on the concentration of leapers.
Huang et al. have also recognized that midair
collisions might sustain grains above the ground reducing the frequency
of collisions with the bed~\cite{Dong,Ren,NHuang}. However, in contrast to our observation, they hypothesized that such a decrease would reduce the mass transport.
Indeed, we confirmed that the mass flux of the no-collision case is below the one of the collisional case ($e=0.7$) even for small $\theta$ close to the transport threshold.
Figure~\ref{concentration}a compares the concentration profiles for $\theta = 0.44$ with and without midair collisions.
The yellow and red (dark) curves are in good agreement with the ones obtained by Jenkins and co-workers with kinetic theory with~\cite{pasini} and without~\cite{jenkins} midair collisions, respectively.

Simultaneously, another important mechanism is activated.
While, without dissipation, particles rebound randomly in air, dissipation tends to align the trajectory of colliding particles~\cite{Duparcmeur}. 
This is expected to occur in the direction of wind. 
We define particle alignment as the variance of the velocity angle with respect to wind direction, i.e., $\sigma^2_\alpha=< \alpha^2>-< \alpha>^2$ with $\alpha = \arctan(v_y/v_x)$. 
A larger variance corresponds to a weaker alignment.
To investigate this, we measure, for the dissipative $(e=0.7)$ and the conservative $(e=1.0)$ cases, the granular temperature $T(y)$ and the inverse of the temperature anisotropy $a(y)=T_y/T_x$ with 
$T_y = <v_y^2>$ and $T_x = <(v_x-<v_x>)^2>$.
The granular temperature in Fig.~\ref{temperature}a displays at every height a larger temperature for $e=1.0$ than for $e =0.7$ with a peak around $y = 50 D_{mean}$.
In the region of the soft-bed, the granular temperature profile is an increasing function of height. This result is in line with what was observed using kinetic theory~\cite{pasini}. 
Figure~\ref{temperature}c shows the dependence of the maximum temperature on $\theta$ for both cases.
The maximum temperature grows with $\theta^{1/2}$ (linear in $u_*$) for $e=0.7$ and increases linearly with $\theta$ (quadratically with $u_*$) for $e = 1.0$, being always larger in the conservative case. 
This higher temperature could be due to a larger dispersion either in magnitude or direction of particle velocities. 
To distinguish these two possibilities we plot, in Fig.~\ref{angle}a, the variance of the velocity angle $\sigma_\alpha^2$  and the inverse of temperature anisotropy $a(y)$.
The plots confirm that a lower temperature in the inelastic case (when compared with the elastic one) corresponds to a lower dispersion in particle velocities, i.e., higher alignment.
Clearly, the alignment of trajectories, which tends to enhance the overall flux, decreases with $e$. 

\begin{figure}[top]
\rotatebox{-90}{\includegraphics[scale=0.34]{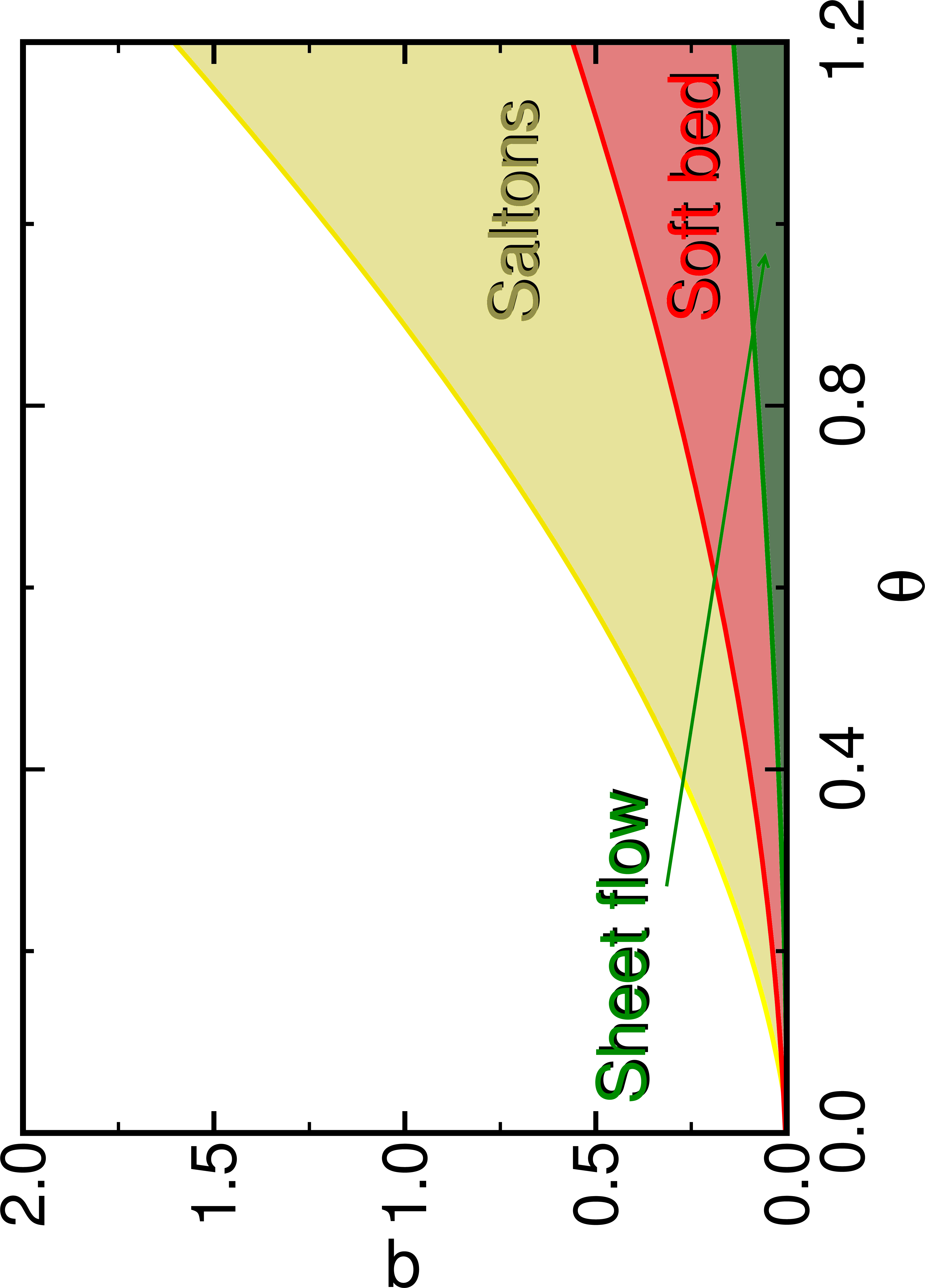}}
\caption{(color online) Contribution to the total saturated flux by sheet flow (lower curve), leapers in the soft-bed (middle curve), and saltons above the soft-bed (upper curve) for $e=0.7$ as function of $\theta$. The green area (lower region) shows the sheet flow in the particle bed $y < h_0$.}
\label{flux}
\end{figure}

In the absence of midair collisions, the distribution of maximum heights can be approximated by a Poisson process and, therefore, is well described by an exponential decay, as shown in Fig.~\ref{angle}b.
With midair collisions, however, this distribution can only be described by the superposition of two exponentials: the first one corresponds to leapers in the simulation without midair collisions, while the second one corresponds to saltons, which typically fly above the leapers. We can now quantitatively define the upper bound of soft-bed as the height above which saltons are in majority.
From the trajectories of individual particles, we can also relate the flying time to the maximum height (inset, Fig.~\ref{angle}b); we observe that saltons stay much longer in the air than leapers.

Three major types of transport contribute to the total flux in saltation: sheet flow, transport of leapers (in the soft-bed), and of saltons (above the soft-bed). 
Once the limits of the soft-bed are identified, we can compute for each $\theta$ the contribution of each mechanism to the total flux, as shown in Fig.~\ref{flux}.
The sheet flow, computed from the mass transport in the particle bed ($y< h_0$), results from the wind shear stress and the creep of particles on the surface. 
The contributions of leapers and saltons are obtained from the fluxes in and above the soft-bed, respectively. 
As observed in the figure, the relative contribution of saltons and leapers significantly increases with $\theta$ in comparison to the sheet flow.
Figure~\ref{flux} also shows that the saltons contribute the most to the total flux.

For the saturated flux, we also reproduced the discontinuity at the onset of saltation reported for $2D$~\cite{jump}.
We also studied the impact of midair collisions in 2D.
The decrease in the spatial dimension enhances the relevance of midair collisions, and the consequences discussed here are even more pronounced.
However, the curves for the saturated flux in $2D$ and $3D$ overlap if $\theta$ in $2D$ is rescaled by an appropriate factor that takes into account the system width.   

Summarizing, the contribution of midair collisions cannot be neglected in saltation, as it would underestimate the mass flux. 
The saltons contribute significantly to the flux enhancement by acquiring large momentum from the wind and using it to eject more particles through splashes.
We provide a new picture of Aeolian saltation, in which the competition between uplift due to midair collisions and alignment due to inelasticity, optimizes the mass transport in the presence of dissipative collisions.
Interestingly, the position of the maximum corresponds to the restitution coefficient typically observed in granular collisions.
Berger et al. have shown a non-trivial dependence of the erosion rate on the restitution coefficient of midair collisions during a lunar landing in the absence of saltation~\cite{Berger}. As a follow-up, continuum models of Aeolian transport would profit from a systematic analysis of the dynamics of particle ejection from the surface in the presence of a soft-bed.

Our results are crucial for future studies since they provide qualitative and quantitative information about the influence of midair collisions, which should be considered in modeling.
Additional work might include the aerodynamic lift of the particles, turbulent wind speed fluctuations, and the electrostatic interaction between particles. 
Further possibilities are the study of the role of collisions in saltation under terrestrial desert wind, subaqueous or Martian conditions. 
It is also still an open question why the maximum granular temperature scales linearly with the shear velocity in the inelastic and quadratically in the elastic case.

\begin{acknowledgments}
This work has been supported by the Brazilian Council for Scientific and
Technological Development CNPq, ETH (Grant No. ETH-10 09-2) and European
Research Council (Grant No. FP7-319968).  We acknowledge also the
financial support from the European Research Council (ERC) Advanced
Grant 319968-FlowCCS and NSFC 41350110226.
\end{acknowledgments}

\bibliography{paper}

\begin{thebibliography}{28}%
\makeatletter
\providecommand \@ifxundefined [1]{%
 \@ifx{#1\undefined}
}%
\providecommand \@ifnum [1]{%
 \ifnum #1\expandafter \@firstoftwo
 \else \expandafter \@secondoftwo
 \fi
}%
\providecommand \@ifx [1]{%
 \ifx #1\expandafter \@firstoftwo
 \else \expandafter \@secondoftwo
 \fi
}%
\providecommand \natexlab [1]{#1}%
\providecommand \enquote  [1]{``#1''}%
\providecommand \bibnamefont  [1]{#1}%
\providecommand \bibfnamefont [1]{#1}%
\providecommand \citenamefont [1]{#1}%
\providecommand \href@noop [0]{\@secondoftwo}%
\providecommand \href [0]{\begingroup \@sanitize@url \@href}%
\providecommand \@href[1]{\@@startlink{#1}\@@href}%
\providecommand \@@href[1]{\endgroup#1\@@endlink}%
\providecommand \@sanitize@url [0]{\catcode `\\12\catcode `\$12\catcode
  `\&12\catcode `\#12\catcode `\^12\catcode `\_12\catcode `\%12\relax}%
\providecommand \@@startlink[1]{}%
\providecommand \@@endlink[0]{}%
\providecommand \url  [0]{\begingroup\@sanitize@url \@url }%
\providecommand \@url [1]{\endgroup\@href {#1}{\urlprefix }}%
\providecommand \urlprefix  [0]{URL }%
\providecommand \Eprint [0]{\href }%
\providecommand \doibase [0]{http://dx.doi.org/}%
\providecommand \selectlanguage [0]{\@gobble}%
\providecommand \bibinfo  [0]{\@secondoftwo}%
\providecommand \bibfield  [0]{\@secondoftwo}%
\providecommand \translation [1]{[#1]}%
\providecommand \BibitemOpen [0]{}%
\providecommand \bibitemStop [0]{}%
\providecommand \bibitemNoStop [0]{.\EOS\space}%
\providecommand \EOS [0]{\spacefactor3000\relax}%
\providecommand \BibitemShut  [1]{\csname bibitem#1\endcsname}%
\let\auto@bib@innerbib\@empty
\bibitem [{\citenamefont {Almeida}\ \emph {et~al.}(2006)\citenamefont
  {Almeida}, \citenamefont {Andrade},\ and\ \citenamefont
  {Herrmann}}]{almeida}%
  \BibitemOpen
  \bibfield  {author} {\bibinfo {author} {\bibfnamefont {M.~P.}\ \bibnamefont
  {Almeida}}, \bibinfo {author} {\bibfnamefont {J.~S.}\ \bibnamefont
  {Andrade}}, \ and\ \bibinfo {author} {\bibfnamefont {H.~J.}\ \bibnamefont
  {Herrmann}},\ }\href@noop {} {\bibfield  {journal} {\bibinfo  {journal}
  {Phys. Rev. Lett.}\ }\textbf {\bibinfo {volume} {96}},\ \bibinfo {pages}
  {018001} (\bibinfo {year} {2006})}\BibitemShut {NoStop}%
\bibitem [{\citenamefont {Almeida}\ \emph {et~al.}(2008)\citenamefont
  {Almeida}, \citenamefont {Parteli}, \citenamefont {Andrade},\ and\
  \citenamefont {Herrmann}}]{Almeida_PNAS}%
  \BibitemOpen
  \bibfield  {author} {\bibinfo {author} {\bibfnamefont {M.~P.}\ \bibnamefont
  {Almeida}}, \bibinfo {author} {\bibfnamefont {E.~J.~R.}\ \bibnamefont
  {Parteli}}, \bibinfo {author} {\bibfnamefont {J.~S.}\ \bibnamefont
  {Andrade}}, \ and\ \bibinfo {author} {\bibfnamefont {H.~J.}\ \bibnamefont
  {Herrmann}},\ }\href {\doibase 10.1073/pnas.0800202105} {\bibfield  {journal}
  {\bibinfo  {journal} {Proc. Natl. Acad. Sci. USA}\ }\textbf {\bibinfo
  {volume} {105}},\ \bibinfo {pages} {6222} (\bibinfo {year}
  {2008})}\BibitemShut {NoStop}%
\bibitem [{\citenamefont {Werner}(1987)}]{werner}%
  \BibitemOpen
  \bibfield  {author} {\bibinfo {author} {\bibfnamefont {B.~T.}\ \bibnamefont
  {Werner}},\ }\href@noop {} {\emph {\bibinfo {title} {A physical model of
  wind-blown sand transport, Ph.D Thesis}}}\ (\bibinfo {address} {Caltec,
  Pasadena},\ \bibinfo {year} {1987})\BibitemShut {NoStop}%
\bibitem [{\citenamefont {McEwan}\ and\ \citenamefont
  {Willetts}(1993)}]{McEwan}%
  \BibitemOpen
  \bibfield  {author} {\bibinfo {author} {\bibfnamefont {I.~K.}\ \bibnamefont
  {McEwan}}\ and\ \bibinfo {author} {\bibfnamefont {B.~B.}\ \bibnamefont
  {Willetts}},\ }\href@noop {} {\bibfield  {journal} {\bibinfo  {journal} {J.
  Fluid Mech.}\ }\textbf {\bibinfo {volume} {252}},\ \bibinfo {pages} {99}
  (\bibinfo {year} {1993})}\BibitemShut {NoStop}%
\bibitem [{\citenamefont {Kok}\ and\ \citenamefont {Renno}(2009)}]{kok}%
  \BibitemOpen
  \bibfield  {author} {\bibinfo {author} {\bibfnamefont {J.~F.}\ \bibnamefont
  {Kok}}\ and\ \bibinfo {author} {\bibfnamefont {N.~O.}\ \bibnamefont
  {Renno}},\ }\href@noop {} {\bibfield  {journal} {\bibinfo  {journal} {J.
  Geophys. Res.}\ }\textbf {\bibinfo {volume} {114}},\ \bibinfo {pages} {149}
  (\bibinfo {year} {2009})}\BibitemShut {NoStop}%
\bibitem [{\citenamefont {P\"ahtz}\ \emph {et~al.}(2012)\citenamefont
  {P\"ahtz}, \citenamefont {Kok},\ and\ \citenamefont {Herrmann}}]{Pahtz}%
  \BibitemOpen
  \bibfield  {author} {\bibinfo {author} {\bibfnamefont {T.}~\bibnamefont
  {P\"ahtz}}, \bibinfo {author} {\bibfnamefont {J.}~\bibnamefont {Kok}}, \ and\
  \bibinfo {author} {\bibfnamefont {H.~J.}\ \bibnamefont {Herrmann}},\
  }\href@noop {} {\bibfield  {journal} {\bibinfo  {journal} {New J. Phys.}\
  }\textbf {\bibinfo {volume} {14}},\ \bibinfo {pages} {43035} (\bibinfo {year}
  {2012})}\BibitemShut {NoStop}%
\bibitem [{\citenamefont {Bagnold}(1941)}]{Bagnold3}%
  \BibitemOpen
  \bibfield  {author} {\bibinfo {author} {\bibfnamefont {R.~A.}\ \bibnamefont
  {Bagnold}},\ }\href@noop {} {\emph {\bibinfo {title} {The Physics of Blown
  Sand and Desert Dunes}}}\ (\bibinfo {address} {Methuen, New York},\ \bibinfo
  {year} {1941})\ p.\ \bibinfo {pages} {265}\BibitemShut {NoStop}%
\bibitem [{\citenamefont {Kok}\ \emph {et~al.}(2012)\citenamefont {Kok},
  \citenamefont {Parteli}, \citenamefont {Michaels},\ and\ \citenamefont
  {Karam}}]{kok-parteli}%
  \BibitemOpen
  \bibfield  {author} {\bibinfo {author} {\bibfnamefont {J.~F.}\ \bibnamefont
  {Kok}}, \bibinfo {author} {\bibfnamefont {E.~R.}\ \bibnamefont {Parteli}},
  \bibinfo {author} {\bibfnamefont {T.}~\bibnamefont {Michaels}}, \ and\
  \bibinfo {author} {\bibfnamefont {D.~B.}\ \bibnamefont {Karam}},\ }\href
  {\doibase doi:10.1088/0034-4885/75/10/106901} {\bibfield  {journal} {\bibinfo
   {journal} {Rep. Prog. Phys.}\ }\textbf {\bibinfo {volume} {75}},\ \bibinfo
  {pages} {106901} (\bibinfo {year} {2012})}\BibitemShut {NoStop}%
\bibitem [{\citenamefont {Williams}(1987)}]{Williams}%
  \BibitemOpen
  \bibfield  {author} {\bibinfo {author} {\bibfnamefont {S.~H.}\ \bibnamefont
  {Williams}},\ }\emph {\bibinfo {title} {A comparative planetological study of
  particle speed and concentration during aeolian saltation}},\ \href@noop {}
  {Ph.D. thesis},\ \bibinfo  {school} {Arizona State University} (\bibinfo
  {year} {1987}),\ \bibinfo {note} {ph.D thesis}\BibitemShut {NoStop}%
\bibitem [{\citenamefont {Soerensen}\ and\ \citenamefont
  {McEwan}(1996)}]{soresen}%
  \BibitemOpen
  \bibfield  {author} {\bibinfo {author} {\bibfnamefont {M.}~\bibnamefont
  {Soerensen}}\ and\ \bibinfo {author} {\bibfnamefont {I.}~\bibnamefont
  {McEwan}},\ }\href@noop {} {\bibfield  {journal} {\bibinfo  {journal}
  {Sedimentology}\ }\textbf {\bibinfo {volume} {43}},\ \bibinfo {pages} {65}
  (\bibinfo {year} {1996})}\BibitemShut {NoStop}%
\bibitem [{\citenamefont {Andreotti}(2004)}]{Andreotti}%
  \BibitemOpen
  \bibfield  {author} {\bibinfo {author} {\bibfnamefont {B.}~\bibnamefont
  {Andreotti}},\ }\href@noop {} {\bibfield  {journal} {\bibinfo  {journal} {J.
  Fluid Mech.}\ }\textbf {\bibinfo {volume} {510}},\ \bibinfo {pages} {47}
  (\bibinfo {year} {2004})}\BibitemShut {NoStop}%
\bibitem [{\citenamefont {L\"ammel}\ \emph {et~al.}(2012)\citenamefont
  {L\"ammel}, \citenamefont {Rings},\ and\ \citenamefont {Kroy}}]{Lammel}%
  \BibitemOpen
  \bibfield  {author} {\bibinfo {author} {\bibfnamefont {M.}~\bibnamefont
  {L\"ammel}}, \bibinfo {author} {\bibfnamefont {D.}~\bibnamefont {Rings}}, \
  and\ \bibinfo {author} {\bibfnamefont {K.}~\bibnamefont {Kroy}},\ }\href
  {\doibase doi:10.1088/1367-2630/14/9/093037} {\bibfield  {journal} {\bibinfo
  {journal} {New J. Phys.}\ }\textbf {\bibinfo {volume} {14}},\ \bibinfo
  {pages} {093037} (\bibinfo {year} {2012})}\BibitemShut {NoStop}%
\bibitem [{\citenamefont {Mitha}\ \emph {et~al.}(1986)\citenamefont {Mitha},
  \citenamefont {Tran}, \citenamefont {Werner},\ and\ \citenamefont
  {Haff}}]{Mitha}%
  \BibitemOpen
  \bibfield  {author} {\bibinfo {author} {\bibfnamefont {S.}~\bibnamefont
  {Mitha}}, \bibinfo {author} {\bibfnamefont {M.~Q.}\ \bibnamefont {Tran}},
  \bibinfo {author} {\bibfnamefont {B.~T.}\ \bibnamefont {Werner}}, \ and\
  \bibinfo {author} {\bibfnamefont {P.~K.}\ \bibnamefont {Haff}},\ }\href@noop
  {} {\bibfield  {journal} {\bibinfo  {journal} {Acta Mechanica}\ }\textbf
  {\bibinfo {volume} {63}},\ \bibinfo {pages} {267} (\bibinfo {year}
  {1986})}\BibitemShut {NoStop}%
\bibitem [{\citenamefont {Anderson}\ and\ \citenamefont
  {Haff}(1988)}]{Anderson}%
  \BibitemOpen
  \bibfield  {author} {\bibinfo {author} {\bibfnamefont {R.~S.}\ \bibnamefont
  {Anderson}}\ and\ \bibinfo {author} {\bibfnamefont {P.~K.}\ \bibnamefont
  {Haff}},\ }\href@noop {} {\bibfield  {journal} {\bibinfo  {journal}
  {Science}\ }\textbf {\bibinfo {volume} {241}},\ \bibinfo {pages} {820 }
  (\bibinfo {year} {1988})}\BibitemShut {NoStop}%
\bibitem [{\citenamefont {Griebel}\ \emph {et~al.}(2007)\citenamefont
  {Griebel}, \citenamefont {Knapek},\ and\ \citenamefont {Zumbusch}}]{Griebel}%
  \BibitemOpen
  \bibfield  {author} {\bibinfo {author} {\bibfnamefont {M.}~\bibnamefont
  {Griebel}}, \bibinfo {author} {\bibfnamefont {S.}~\bibnamefont {Knapek}}, \
  and\ \bibinfo {author} {\bibfnamefont {G.}~\bibnamefont {Zumbusch}},\
  }\href@noop {} {\emph {\bibinfo {title} {Numerical Simulation in Molecular
  Dynamics: Numerics, Algorithms, Parallelization, Applications}}}\ (\bibinfo
  {publisher} {Springer},\ \bibinfo {address} {New York},\ \bibinfo {year}
  {2007})\BibitemShut {NoStop}%
\bibitem [{\citenamefont {Carneiro}\ \emph {et~al.}(2011)\citenamefont
  {Carneiro}, \citenamefont {P\"ahtz},\ and\ \citenamefont {Herrmann}}]{jump}%
  \BibitemOpen
  \bibfield  {author} {\bibinfo {author} {\bibfnamefont {M.~V.}\ \bibnamefont
  {Carneiro}}, \bibinfo {author} {\bibfnamefont {T.}~\bibnamefont {P\"ahtz}}, \
  and\ \bibinfo {author} {\bibfnamefont {H.~J.}\ \bibnamefont {Herrmann}},\
  }\href@noop {} {\bibfield  {journal} {\bibinfo  {journal} {Phys. Rev. Lett.}\
  }\textbf {\bibinfo {volume} {107}},\ \bibinfo {pages} {098001} (\bibinfo
  {year} {2011})}\BibitemShut {NoStop}%
\bibitem [{SM()}]{SM}%
  \BibitemOpen
  \href {...} {}\bibinfo {note} {See Supplemental Material at}\BibitemShut
  {NoStop}%
\bibitem [{\citenamefont {Rioual}\ \emph {et~al.}(2000)\citenamefont {Rioual},
  \citenamefont {Valance},\ and\ \citenamefont {Bideau}}]{rioual}%
  \BibitemOpen
  \bibfield  {author} {\bibinfo {author} {\bibfnamefont {F.}~\bibnamefont
  {Rioual}}, \bibinfo {author} {\bibfnamefont {A.}~\bibnamefont {Valance}}, \
  and\ \bibinfo {author} {\bibfnamefont {D.}~\bibnamefont {Bideau}},\
  }\href@noop {} {\bibfield  {journal} {\bibinfo  {journal} {Phys. Rev. E}\
  }\textbf {\bibinfo {volume} {62}},\ \bibinfo {pages} {2450} (\bibinfo {year}
  {2000})}\BibitemShut {NoStop}%
\bibitem [{\citenamefont {Rioual}(2002)}]{rioual_thesis}%
  \BibitemOpen
  \bibfield  {author} {\bibinfo {author} {\bibfnamefont {F.}~\bibnamefont
  {Rioual}},\ }\href@noop {} {Ph.D. thesis},\ \bibinfo  {school} {University of
  Rennes 1} (\bibinfo {year} {2002})\BibitemShut {NoStop}%
\bibitem [{\citenamefont {Rioual}\ \emph {et~al.}(2003)\citenamefont {Rioual},
  \citenamefont {Valance},\ and\ \citenamefont {Bideau}}]{rioual_2003}%
  \BibitemOpen
  \bibfield  {author} {\bibinfo {author} {\bibfnamefont {F.}~\bibnamefont
  {Rioual}}, \bibinfo {author} {\bibfnamefont {A.}~\bibnamefont {Valance}}, \
  and\ \bibinfo {author} {\bibfnamefont {D.}~\bibnamefont {Bideau}},\
  }\href@noop {} {\bibfield  {journal} {\bibinfo  {journal} {Europhys. Lett.}\
  }\textbf {\bibinfo {volume} {61}},\ \bibinfo {pages} {194} (\bibinfo {year}
  {2003})}\BibitemShut {NoStop}%
\bibitem [{\citenamefont {Banks}\ \emph {et~al.}(2005)\citenamefont {Banks},
  \citenamefont {Bridges},\ and\ \citenamefont {Benzit}}]{nasa}%
  \BibitemOpen
  \bibfield  {author} {\bibinfo {author} {\bibfnamefont {M.}~\bibnamefont
  {Banks}}, \bibinfo {author} {\bibfnamefont {N.}~\bibnamefont {Bridges}}, \
  and\ \bibinfo {author} {\bibfnamefont {M.}~\bibnamefont {Benzit}},\
  }\href@noop {} {\bibfield  {journal} {\bibinfo  {journal} {Proceedings of the
  36th Annual Lunar and Planetary Science Conference}\ ,\ \bibinfo {pages}
  {2116}} (\bibinfo {year} {2005})}\BibitemShut {NoStop}%
\bibitem [{\citenamefont {Dong}\ \emph {et~al.}(2005)\citenamefont {Dong},
  \citenamefont {Huang},\ and\ \citenamefont {Liu}}]{Dong}%
  \BibitemOpen
  \bibfield  {author} {\bibinfo {author} {\bibfnamefont {Z.}~\bibnamefont
  {Dong}}, \bibinfo {author} {\bibfnamefont {N.}~\bibnamefont {Huang}}, \ and\
  \bibinfo {author} {\bibfnamefont {X.}~\bibnamefont {Liu}},\ }\href@noop {}
  {\bibfield  {journal} {\bibinfo  {journal} {J. Geophys. Res.}\ }\textbf
  {\bibinfo {volume} {110}},\ \bibinfo {pages} {D24113} (\bibinfo {year}
  {2005})}\BibitemShut {NoStop}%
\bibitem [{\citenamefont {Ren}\ and\ \citenamefont {Huang}(2010)}]{Ren}%
  \BibitemOpen
  \bibfield  {author} {\bibinfo {author} {\bibfnamefont {S.}~\bibnamefont
  {Ren}}\ and\ \bibinfo {author} {\bibfnamefont {N.}~\bibnamefont {Huang}},\
  }\href@noop {} {\bibfield  {journal} {\bibinfo  {journal} {Eur. Phys. J. E}\
  }\textbf {\bibinfo {volume} {33}},\ \bibinfo {pages} {351} (\bibinfo {year}
  {2010})}\BibitemShut {NoStop}%
\bibitem [{\citenamefont {Huang}\ \emph {et~al.}(2007)\citenamefont {Huang},
  \citenamefont {Zhang},\ and\ \citenamefont {D'Adamo}}]{NHuang}%
  \BibitemOpen
  \bibfield  {author} {\bibinfo {author} {\bibfnamefont {N.}~\bibnamefont
  {Huang}}, \bibinfo {author} {\bibfnamefont {Y.}~\bibnamefont {Zhang}}, \ and\
  \bibinfo {author} {\bibfnamefont {R.}~\bibnamefont {D'Adamo}},\ }\href@noop
  {} {\bibfield  {journal} {\bibinfo  {journal} {J. Geophys. Res.}\ }\textbf
  {\bibinfo {volume} {112}},\ \bibinfo {pages} {D08206} (\bibinfo {year}
  {2007})}\BibitemShut {NoStop}%
\bibitem [{\citenamefont {Pasini}\ and\ \citenamefont
  {Jenkins}(2005)}]{pasini}%
  \BibitemOpen
  \bibfield  {author} {\bibinfo {author} {\bibfnamefont {J.~M.}\ \bibnamefont
  {Pasini}}\ and\ \bibinfo {author} {\bibfnamefont {J.~T.}\ \bibnamefont
  {Jenkins}},\ }\href@noop {} {\bibfield  {journal} {\bibinfo  {journal} {Phil.
  Trans. R. Soc. A}\ }\textbf {\bibinfo {volume} {363}},\ \bibinfo {pages}
  {1625} (\bibinfo {year} {2005})}\BibitemShut {NoStop}%
\bibitem [{\citenamefont {Jenkins}\ \emph {et~al.}(2010)\citenamefont
  {Jenkins}, \citenamefont {Cantat},\ and\ \citenamefont {Valance}}]{jenkins}%
  \BibitemOpen
  \bibfield  {author} {\bibinfo {author} {\bibfnamefont {J.~T.}\ \bibnamefont
  {Jenkins}}, \bibinfo {author} {\bibfnamefont {I.}~\bibnamefont {Cantat}}, \
  and\ \bibinfo {author} {\bibfnamefont {A.}~\bibnamefont {Valance}},\
  }\href@noop {} {\bibfield  {journal} {\bibinfo  {journal} {Phys. Rev. E}\
  }\textbf {\bibinfo {volume} {82}},\ \bibinfo {pages} {020301} (\bibinfo
  {year} {2010})}\BibitemShut {NoStop}%
\bibitem [{\citenamefont {Duparcmeur}\ \emph {et~al.}(1995)\citenamefont
  {Duparcmeur}, \citenamefont {Herrmann},\ and\ \citenamefont
  {Troadec}}]{Duparcmeur}%
  \BibitemOpen
  \bibfield  {author} {\bibinfo {author} {\bibfnamefont {Y.~L.}\ \bibnamefont
  {Duparcmeur}}, \bibinfo {author} {\bibfnamefont {H.}~\bibnamefont
  {Herrmann}}, \ and\ \bibinfo {author} {\bibfnamefont {J.}~\bibnamefont
  {Troadec}},\ }\href@noop {} {\bibfield  {journal} {\bibinfo  {journal} {J.
  Physique I}\ }\textbf {\bibinfo {volume} {5}},\ \bibinfo {pages} {1119 }
  (\bibinfo {year} {1995})}\BibitemShut {NoStop}%
\bibitem [{\citenamefont {Berger}\ \emph {et~al.}(2013)\citenamefont {Berger},
  \citenamefont {Anand}, \citenamefont {Metzger},\ and\ \citenamefont
  {Hrenya}}]{Berger}%
  \BibitemOpen
  \bibfield  {author} {\bibinfo {author} {\bibfnamefont {K.~J.}\ \bibnamefont
  {Berger}}, \bibinfo {author} {\bibfnamefont {A.}~\bibnamefont {Anand}},
  \bibinfo {author} {\bibfnamefont {P.~T.}\ \bibnamefont {Metzger}}, \ and\
  \bibinfo {author} {\bibfnamefont {C.~M.}\ \bibnamefont {Hrenya}},\
  }\href@noop {} {\bibfield  {journal} {\bibinfo  {journal} {Phys. Rev. E}\
  }\textbf {\bibinfo {volume} {87}},\ \bibinfo {pages} {022205} (\bibinfo
  {year} {2013})}\BibitemShut {NoStop}%
\end{thebibliography}%



\section*{Supplementary material}

# Supplemental Material: Mid-air collisions enhance saltation

M. V. Carneiro[1], N. A. M. Araújo[1], T. Pähtz[2,3], and H. J. Herrmann[1,4]

[1]*Institut für Baustoffe, ETH-Hönggerberg, Schafmattstrasse 6, 8093 Zürich, Switzerland*
[2]*Department of Ocean Science and Engineering, Zhejiang University, 310058 Hangzhou, China*
[3]*State Key Laboratory of Satellite Ocean Environment Dynamics,
Second Institute of Oceanography, Hangzhou, China and*
[4]*Departamento de Física, Universidade Federal do Ceará, 60451-970 Fortaleza, Ceará, Brazil*

(Dated: December 18, 2012)

## I. PARTICE DYNAMICS

Trajectories and velocities of the particles are obtained iteratively by solving the Newton equations of motion, through the velocity-Störmer-Verlet scheme [1]. The contact between particles is described by the spring dashpot potential, which has an elastic and a dissipative contribution acting on each particle. When two particles $i$ and $j$ overlap (i.e. when their distance is smaller than the sum of their radii) an elastic force is applied,

$$\mathbf{F}_{el}^{(i)} = km_i[|\mathbf{r}_{ij}| - 1/2(d_i + d_j)]\frac{\mathbf{r}_{ij}}{|\mathbf{r}_{ij}|}, \quad (1)$$

where $k = 0.5$ is a spring constant, $d_i$ and $d_j$ are the diameters, $m_i$ is the mass of particle $i$, and $\mathbf{r}_{ij}$ points from particle $i$ to $j$. A dissipation force is also applied accounting for the inelasticity of the collision,

$$\mathbf{F}_{diss}^{(i)} = -\gamma \mathbf{v}_{ij}, \quad (2)$$

where $\mathbf{v}_{ij} = \mathbf{v}_i - \mathbf{v}_j$ is the relative velocity and $\gamma$ is the dissipation coefficient. The coefficient of restitution $e$ is given by the ratio between the absolute velocities after and before the collision. For particle-wall collisions, the same forces act as if particle $i$ would collide with another particle of diameter $d_i$ [2]. For simplicity, friction and rotation are neglected.

## II. THE WIND PROFILE

A logarithmic velocity field mimics the wind profile in the absence of saltating grains in the $x$-direction with

$$u(y) = \frac{u_*}{\kappa} \ln \frac{y - h_0}{y_0}, \quad (3)$$

where $y_0 = D_{mean}/30$ is the roughness of the bed with $D_{mean}$ the mean diameter of the particles, $h_0$ the bed height, $\kappa = 0.4$ the von Kármán constant, and $u_*$ the wind shear velocity. In the presence of grains, the wind strength is substantially reduced due to the momentum transferred to the grains [3]. The grain stress $\tau_g(y)$ quantifies the average horizontal force per unit area $f$ that the wind applies on the grains above $y$ [4], i.e.,

$$\tau_g(y) = \int_y^\infty f(y')dy'. \quad (4)$$

where $f$ is the average horizontal force per unit volume. The modified wind shear velocity $u_\tau(y)$ is the fluid stress left at the height $y$ after the momentum transfer,

$$u_\tau(y) = u_*\sqrt{1 - \frac{\tau_g}{\rho_w u_*^2}}, \quad (5)$$

where $\rho_w$ is the air density. To obtain the modified wind profile [5], we solve,

$$\frac{du}{dy} = \frac{u_\tau(y)}{\kappa y}, \quad (6)$$

considering $u_\tau(y)$ constant within an interval $dy$.

The numerical solution is achieved iteratively. The position of the bed surface is used as starting point of the integration of the wind profile. However, the particle splash and the sheet flow dynamically affect the shape of the bed surface. Consequently, we need to find $h_0$ for every time step. We use a criterion based on the wind values to define an approximate position for the particle surface. High density areas strongly reduce wind velocities. If the calculated velocity $v_i$ in area $y_i$ is below $0.1u_*$, we assume that this area contains the particle bed and the velocity is set to zero. This means that $h_0$ is chosen to be the point where the calculated velocity exceeds $0.1u_*$.

The wind drag is the only external force applied to the particles along the $x$-direction,

$$\mathbf{F}_d = -\frac{\pi D^2}{8}\rho_w C_d v_r \mathbf{v_r}, \quad (7)$$

where $\rho_w$ is the air density and $\mathbf{v_r} = \mathbf{v} - \mathbf{u}$ is the velocity difference between particle and wind, with $v_r = |\mathbf{v_r}|$. The drag coefficient $C_d$ proposed by Cheng [6] is suited to model grains with irregular, natural, shapes, and is given by,

$$C_d = \left[\left(\frac{32}{Re}\right)^{2/3} + 1\right]^{3/2}, Re = \frac{\rho_w v_r D_{mean}}{\mu}, \quad (8)$$

where $\mu = 1.8702 \times 10^{-5}$ kg/(m.s) is the dynamic viscosity and $Re$, the Reynolds number.

[1] M. Griebel, S. Knapek, and G. Zumbusch, *Numerical Simulation in Molecular Dynamics: Numerics, Algorithms, Parallelization, Applications* (Springer, 2007).
[2] H. Herrmann, Physica A **191**, 236 (1992).
[3] R. S. Anderson and P. K. Haff, Science **241(4867)**, 820 (1988).
[4] J. E. Ungar and P. Haff, Sedimentology **34(2)**, 289 (1987).
[5] R. S. Anderson and P. K. Haff, Acta Mech. **1**, 21 (1991).
[6] N. S. Cheng, J. Hydraul. Eng. **123**, 149 (1997).



\end{document}